# Improved fibre dispersion estimation using b-tensor encoding

*Michiel Cottaar[1,@], Filip Szczepankiewicz[2,3,4] , Matteo Bastiani[5,6,1], Moises Hernandez-Fernandez[1,7], Stamatios N. Sotiropoulos[5,6,1], Markus Nilsson[4] , Saad Jbabdi[1]*

[1] Wellcome Centre for Integrative Neuroimaging (WIN) - Centre for Functional Magnetic Resonance Imaging of the Brain (FMRIB), University of Oxford, UK
   @ Corresponding author; email: Michiel.Cottaar@ndcn.ox.ac.uk
   [2] Harvard Medical School, Boston, MA, USA
   [3] Radiology, Brigham and Women's Hospital, Boston, MA, USA
   [4] Clinical Sciences Lund, Lund University, Lund, Sweden
   [5] Sir Peter Mansfield Imaging Centre, School of Medicine, University of Nottingham, UK
   [6] NIHR Biomedical Research Centre, University of Nottingham, UK
   [7] NVIDIA, Santa Clara, CA, USA

## *Abstract*

Measuring fibre dispersion in white matter with diffusion magnetic resonance imaging (MRI) is limited by an inherent degeneracy between fibre dispersion and microscopic diffusion anisotropy (i.e., the diffusion anisotropy expected for a single fibre orientation). This means that estimates of fibre dispersion rely on strong assumptions, such as constant microscopic anisotropy throughout the white matter or specific biophysical models. Here we present a simple approach for resolving this degeneracy using measurements that combine linear (conventional) and spherical tensor diffusion encoding.

To test the accuracy of the fibre dispersion when our microstructural model is only an approximation of the true tissue structure, we simulate multi-compartment data and fit this with a single-compartment model. For such overly simplistic tissue assumptions, we show that the bias in fibre dispersion is greatly reduced (~5x) for single-shell linear and spherical tensor encoding data compared with single-shell or multi-shell conventional data. In in-vivo data we find a consistent estimate of fibre dispersion as we reduce the b-value from 3 to 1.5 ms/µm$^2$, increase the repetition time, increase the echo time, or increase the diffusion time. We conclude that the addition of spherical tensor encoded data to conventional linear tensor encoding data greatly reduces the sensitivity of the estimated fibre dispersion to the model assumptions of the tissue microstructure.

## *Introduction*

Diffusion MRI is commonly used to reconstruct in-vivo white matter tracts and estimate connectivity between brain regions. This requires an estimation of one or more fibre orientations in every white matter voxel. A wide variety of methods have been proposed to deconvolve the diffusion MRI signal to extract these main fibre orientations (Basser et al., 2000; Tuch, 2004; Anderson, 2005; Behrens et al., 2007; Tournier et al., 2007; Descoteaux et al., 2007; Dell'Acqua et al., 2007, 2010). While these approaches can disagree on the number of crossing fibre populations, the fibre orientations tend to be in good agreement with each other as well as with fibre orientations estimated from histology (Sotiropoulos et al., 2013; Seehaus et al., 2015; Schilling et al., 2016; Salo et al., 2018).

A full characterization of the fibre orientation distribution function (fODF) does not only require an estimate of the mean orientation of each fibre, but also the dispersion of fibre orientations around the mean orientation. Several approaches to measure fibre dispersion have been proposed (Kaden et al., 2007; Savadjiev et al., 2008; Sotiropoulos et al., 2012; Tariq et al., 2016; Zhang et al., 2012). Most are based on spherical deconvolution (Dell'Acqua and Tournier, 2018), where the diffusion signal $S$ is modelled as the convolution between the fODF and a single-fibre response function $R$:

$$S = R * \text{fODF} + S_{\text{other}}, \tag{1}$$

where $S_{\text{other}}$ represents the signal contribution from other compartments not described by the fODF (e.g., partial volume due to free water or cerebrospinal fluid). Measuring fibre dispersion (i.e., the width of the fODF) using this approach requires to overcome the inherent degeneracy between the width of the response function $R$ and the width of the fODF. A more isotropic signal could be explained by either an increase in fibre dispersion or a decrease in anisotropy of the response function. Multiple approaches have been proposed to break this degeneracy, such as assuming a constant response function throughout the brain as in constrained spherical deconvolution (Tournier et al., 2007, 2004), assuming that the response function can be described by a diffusion tensor giving a constant anisotropy across b-values (Kaden et al., 2007, 2016b; Sotiropoulos et al., 2012), or assuming specific biophysical models for the width of the response function and the signal from other compartments as in NODDI (Zhang et al., 2012). The latter two strategies break the degeneracy by acquiring diffusion data at multiple b-values and making assumptions on how the width of the response function varies with b-value. While this does break the degeneracy, the accuracy of the resulting fibre dispersion will depend on the accuracy of the assumptions.

We propose to use b-tensor encoding (Westin et al., 2016) to resolve the degeneracy between the width of the response function and that of the fODF. Our approach yields accurate estimates of fibre dispersion that are only weakly dependent on a priori assumptions.

Our method combines data from the standard Stejskal-Tanner sequence (Stejskal and Tanner, 1965), which is sensitive to diffusion along one direction (i.e., linear tensor encoding) with data sensitive to diffusion in all directions (i.e., spherical tensor encoding) at the same b-value, echo time, and repetition time. Previous studies have shown that combining data acquired with at least two shapes of the b-tensor allows for the measurement of microscopic anisotropy, which characterises the microscopic anisotropy unaffected by orientation dispersion or crossing fibres (Jespersen et al., 2013; Lasič et al., 2014; Shemesh et al., 2015; Szczepankiewicz et al., 2015). This adds an additional constraint to resolve a degeneracy in biophysical models of white matter microstructure (Lampinen et al., 2017, 2019; Coelho et al., 2019; Reisert et al., 2019). Here we investigate whether spherical tensor diffusion encoding provides sufficient information to improve deconvolution of the diffusion MRI signal and retrieve fibre dispersion. By comparing the observed macroscopic diffusion anisotropy (e.g., FA) with that expected from the microscopic anisotropy, an "order parameter", which is sensitive to the alignment of fibre orientations within a voxel, can be measured (Lasič et al., 2014; Szczepankiewicz et al., 2015). How accurately this parameter describes fibre dispersion in a voxel has not been investigated yet.

First, we present the theory for how the combination of linear and spherical tensor encoding provides an fODF-independent measure of microscopic anisotropy in a voxel. We then present a single-compartment model of fibre dispersion in a voxel that can be fitted to data acquired with just linear or linear and spherical encoding. Although this model is highly simplified, we show it still gives an accurate measure of fibre dispersion for a single shell of linear tensor and spherical tensor encoded data in a simulated voxel containing multiple compartments. Because ground-truth fibre dispersion is unknown in vivo, we cannot directly test the accuracy our fibre dispersion estimate. Instead, we evaluate our model on in-vivo data by investigating the consistency of the fibre dispersions across b-values and echo times.

## *Theory*

### *Microscopic anisotropy from the spherical mean*

In the Stejskal-Tanner sequence (Stejskal and Tanner, 1965), diffusion encoding is obtained by separating two equivalent gradient pulses by a 180-degree refocussing pulse, which sensitizes the signal to diffusion along the gradient direction $\hat{g}$. For a single compartment with Gaussian diffusion characterised by a diffusion tensor **D** symmetric around the compartment orientation $\hat{v}$ with eigenvalues $\lambda_1 = d_\parallel$, $\lambda_2 = \lambda_3 = d_\perp$ this leads to a signal attenuation given by (Basser et al., 1994):

$$S_{\text{linear}} = S_0 e^{-b\,\hat{g}^T \cdot \mathbf{D} \cdot \hat{g}} = S_0 e^{-b\,d_\perp} e^{-b(d_\parallel - d_\perp)(\hat{g}\cdot\hat{v})^2}, \tag{2}$$

When averaged across sufficient gradient orientations (Li et al., 2018; Szczepankiewicz et al., 2016b) sampled uniformly across the unit sphere this leads to a spherical mean signal of (Lindblom et al., 1977; Callaghan et al., 1979; Jespersen et al., 2013; Lasič et al., 2014):

$$\langle S_{\text{linear}} \rangle = S_0 e^{-b\, d_\perp} \sqrt{\frac{\pi}{4b(d_\parallel - d_\perp)}} \, \text{erf}\left(\sqrt{b(d_\parallel - d_\perp)}\right). \tag{3}$$

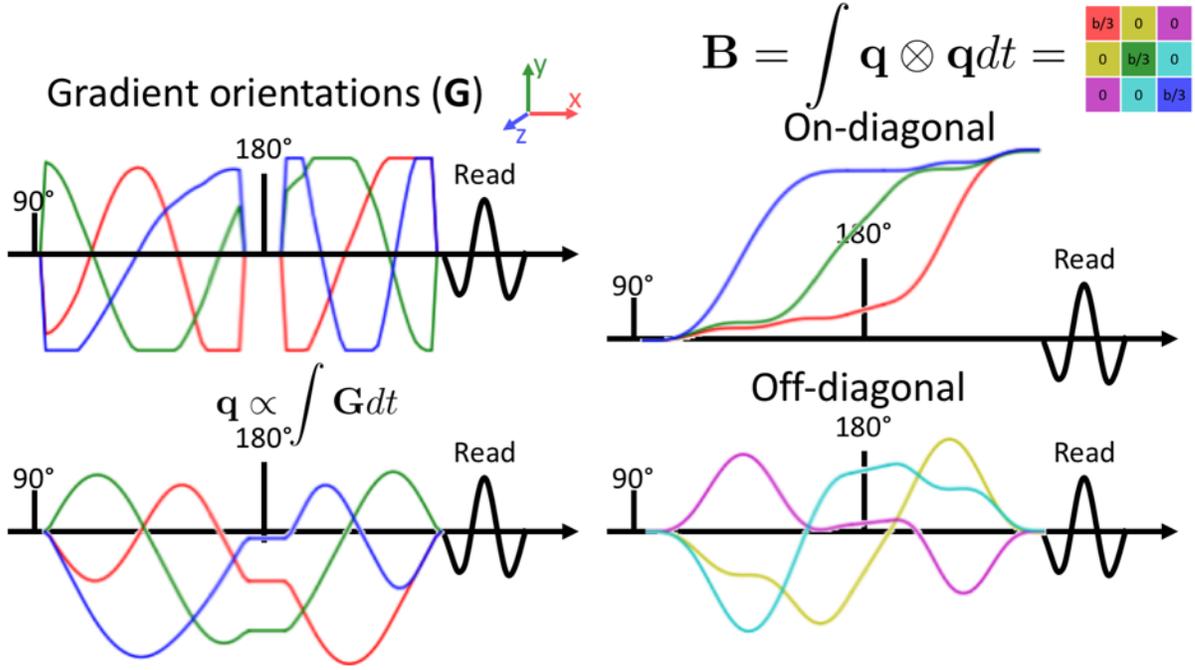

**Figure 1** The gradient waveforms (upper left) adopted in this work to achieve an isotropic sensitivity to diffusion. The resulting path through **q**-space is shown in the lower left and the build-up of sensitivity to the diffusion tensor (i.e., the **B**-tensor) is shown on the right (on- and off-diagonal elements are plotted separately using the colour coding shown in the tensor in the upper right). The gradient waveforms have been designed to obtain a **B**-tensor that is a multiple of the unit tensor (right) and are corrected for the bias that concomitant gradients might cause in such asymmetric gradient waveforms (Szczepankiewicz et al., 2019b).

For an accurate estimate of fibre dispersion we combine the signal sensitive to diffusion along a single direction described above with a signal that is equally sensitive to diffusion along all directions (Mori and Zijl, 1995; Wong et al., 1995). This is attained by altering the gradient waveforms to follow a **q**-space trajectory (Eriksson et al., 2013; Westin et al., 2014; Sjölund et al., 2015; Westin et al., 2016), which under the assumption that the diffusion can be described as a mixture of Gaussians leads to an isotropic sensitivity to diffusion (Figure 1). We refer to the resulting signal as spherical tensor encoded data and signal sensitive to diffusion in a single direction as linear tensor encoded data (Westin et al., 2016).

For spherical tensor encoding, the equal sensitivity to diffusion along all directions ensures that the signal attenuation in each compartment can be described by the isotropic diffusion in that compartment ($d_{iso} = \frac{1}{3}(d_\| + 2d_\perp)$):

$$\langle S_{spherical} \rangle = S_{spherical} = S_0 e^{-b d_{iso}}, \tag{4}$$

where the b-value is given by the trace of the **B**-tensor (Figure 1).

By dividing eq. 3 by eq. 4 we find that the ratio of the spherical mean of the linear tensor encoded signal over the spherical tensor encoded signal is determined only by the anisotropy of the axisymmetric diffusion tensor as measured by $b(d_\| - d_\perp)$:

$$\frac{\langle S_{linear} \rangle}{S_{spherical}} = e^{\frac{1}{3}b(d_\| - d_\perp)} \sqrt{\frac{\pi}{4b(d_\| - d_\perp)}} \operatorname{erf}(\sqrt{b(d_\| - d_\perp)}). \tag{5}$$

This equation can be used to estimate the microscopic anisotropy, which we will define in this work as $(d_\| - d_\perp)$. This is closely related to the microscopic anisotropy defined by eq. 1 in Shemesh et al., (2015), which for an axisymmetric tensor becomes $\mu A = \frac{2}{3}(d_\| - d_\perp)^2$.

This signal ratio is equal to one for a compartment with isotropic diffusion ($d_{iso} = d_\| = d_\perp$) and increases as the diffusion anisotropy increases (Figure 2). Importantly, this ratio only relies on the spherical mean of the diffusion signal and hence provides an independent measure of the diffusion anisotropy from the signal anisotropy usually measured as the fractional anisotropy (FA) (Kaden et al., 2016a; F Szczepankiewicz et al., 2016).

The above formulation assumes a description in terms of a single diffusion tensor, but the complexities of brain tissue may be better modelled using multiple compartments to represent axons and dendrites with various orientations, extra-axonal space, cell bodies (including neurons and glia), and cerebrospinal fluid (CSF). Keeping the assumption that each of these compartments can be described by an axisymmetric diffusion tensor, the signal ratio is given by:

$$\frac{\langle S_{\text{linear}} \rangle}{S_{\text{spherical}}} = \sum_i f_i e^{\frac{1}{3}b(d_\parallel - d_\perp)_i} \sqrt{\frac{\pi}{4b(d_\parallel - d_\perp)_i}} \, \text{erf}(\sqrt{b(d_\parallel - d_\perp)_i}), \tag{6}$$

where $f_i$ is a term describing the b-value weighted signal fraction of each compartment $i$, which depends on the sequence's *b*-value, echo time and repetition time:

$$f_i = \frac{S_{0,i} e^{-b d_{\text{iso},i}}}{\sum_j S_{0,j} e^{-b d_{\text{iso},j}}} \tag{7}$$

Hence, the microscopic anisotropy (i.e., $d_\parallel - d_\perp$) estimated by inverting eq. 5 would be expected to give an unbiased estimate of the true microscopic anisotropy if all compartments have the same microscopic anisotropy, but only differed in their orientation (e.g., dispersing fibres with no extra-axonal contribution). For multiple compartments with different microscopic anisotropy, inverting eq. 5 would give a mean microscopic anisotropy in these compartments weighted by their relative signal fractions.

The signal ratio in eq. 5 can be closely approximated by its second-order Taylor expansion (Jespersen et al., 2013):

$$\frac{\langle S_{\text{linear}} \rangle}{S_{\text{spherical}}} \approx 1 + \frac{2}{45} b^2 (d_\parallel - d_\perp)^2. \tag{8}$$

This approximation holds up to $b(d_\parallel - d_\perp) \approx 6$ (Figure 2). Adopting this approximation across multiple compartments (eq. 6), it can be shown a ground truth value for the microscopic anisotropy estimated from eq. 5 can be estimated using (Ianuş et al., 2018):

$$(d_\parallel - d_\perp)^2_{\text{fit}} \approx \sum_i f_i (d_\parallel - d_\perp)_i^2 \quad (9)$$

Importantly, this estimate of the microscopic anisotropy is obtained by combining linear and spherical tensor encoding at a single b-value and hence does not rely on the assumption that this micro-anisotropy does not change as a function of b-value (Kaden et al., 2016b). In the remainder of this work we show that this weighted mean gives a good approximation of the width of the response function needed to deconvolve the diffusion signal to obtain a measure of fibre dispersion.

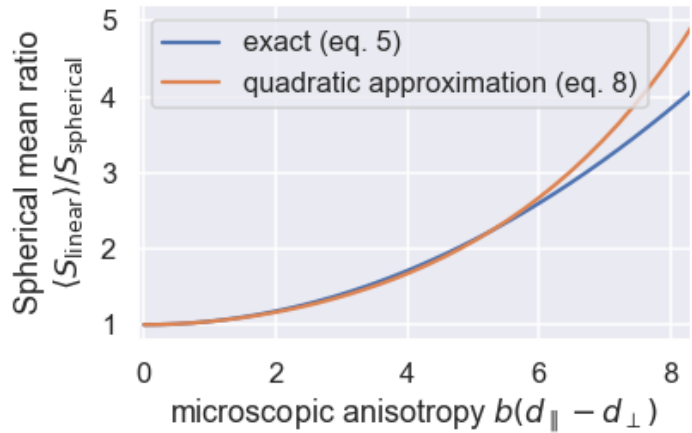

**Figure 2** Dependence of the ratio of the spherical mean of the linear tensor encoded signal (eq. 5) over the spherical tensor encoded signal on the microscopic anisotropy multiplied by the b-value (blue). As the microscopic anisotropy (or b-value) increases the ratio increases from a ratio of one for an isotropic medium. The second-order Taylor expansion (eq. 8) has been overlaid in orange.

## *Deconvolving the diffusion signal*

To investigate the reliability of the fibre dispersion derived by combining linear tensor and spherical tensor encoded diffusion data, we utilize a single-compartment model of dispersing zeppelins (i.e., prolate axisymmetric diffusion tensors). For such a set of compartments with identical diffusivities, but a range of orientations ($\hat{v}$) described by the fODF, the linear tensor encoded signal is given by:

$$S_{\text{linear}} = S_0 e^{-bd_\perp} \int \text{fODF}(v) e^{-b\,(d_\parallel - d_\perp)\,(\hat{v}\cdot\hat{g})^2} d\hat{v} \qquad (10)$$

We will assume that the fODF can be described by a Bingham distribution characterised by the Bingham matrix **Z**:

$$fODF(\hat{v}) = \frac{1}{{}_1F_1\left(\frac{1}{2};\frac{3}{2};\mathbf{Z}\right)} e^{-\hat{v}^T \cdot \mathbf{Z} \cdot \hat{v}}, \qquad (11)$$

$$\text{with } \mathbf{Z} = \mathbf{R} \cdot \begin{bmatrix} 0 & 0 & 0 \\ 0 & k_1 & 0 \\ 0 & 0 & k_2 \end{bmatrix} \cdot \mathbf{R}^T, \qquad (12)$$

where ${}_1F_1$ is a hypergeometric function with a matrix argument and **R** is a rotation matrix (Sotiropoulos et al., 2012). The maximum of this fODF is along the x-axis rotated by **R**, with the dispersion along the rotated y- and z-axis described by $k_1$ and $k_2$ respectively. A higher $k_1$ or $k_2$ corresponds to a smaller dispersion along that axis through a non-linear relationship (Sotiropoulos et al., 2012; Tariq et al., 2016). Although $k_1$ and $k_2$ are convenient in fitting, for ease of interpretation we will instead report the angle containing 50% of fibres along the major and minor axes of dispersion in this work. For our purposes in this work, modelling the fODF as a Bingham distribution rather than using the commonly adopted spherical harmonics has the advantage that the Bingham distribution explicitly includes two parameters representing the dispersion (i.e., $k_1$ and $k_2$).

Substituting eq. 11 in eq. 10 and solving the integral gives the dispersing zeppelin model for linear tensor encoding (we also substitute $d_\perp$ in the exponent with $d_{iso} - (d_\parallel - d_\perp)/3$):

$$S_{\text{linear}} = S_0 e^{-b\, d_{iso}} e^{b(d_\parallel - d_\perp)/3}\, \frac{{}_1F_1\left(\frac{1}{2}, \frac{3}{2}; \mathbf{Z} - b(d_\parallel - d_\perp)\hat{g}\cdot\hat{g}^T\right)}{{}_1F_1\left(\frac{1}{2}, \frac{3}{2}; \mathbf{Z}\right)}. \tag{13}$$

We approximate the hypergeometric function numerically using the approach described in Kume and Wood (2005). The spherical tensor encoded signal is independent of the fODF and given by eq. 4. Similarly, the ratio $\langle S_{\text{linear}}\rangle/S_{\text{spherical}}$ is independent of the fODF and given by eq. 5. The free parameters in this model are the signal amplitude at b=0 ($S_0$), the isotropic diffusivity ($d_{iso}$), the microscopic anisotropy ($d_\parallel - d_\perp$), the orientation of the Bingham matrix (**R**) and the dispersion parameters $k_1$ and $k_2$, as encoded in the Bingham matrix (eq. 12). When fitting to single-shell data, two of the parameters were merged into a single parameter: the isotropic diffusion-weighted signal amplitude ($S_{\text{dw}} = S_0 e^{-b\, d_{iso}}$).

Our main focus here will be on the fibre dispersion estimates $k_1$ and $k_2$, for which we shall show that a single shell of diffusion data is sufficient as long as it contains both linear and spherical tensor encoding. This is plausible as the spherical tensor encoding provides a direct estimate of $S_{\text{dw}}$ and the ratio of the signal from the spherical mean of the linear tensor encoding and the spherical tensor encoding provides an estimate of the microscopic anisotropy ($d_\parallel - d_\perp$, eq. 5, Figure 2). Hence, the angular contrast in the linear tensor encoded data only has to constrain the Bingham matrix (i.e., the fibre orientation and dispersion).

## *Methods*

### *Simulating data*

We simulate diffusion data that do not match the assumptions made in our model in order to test the robustness of the fibre dispersion that can be estimated by combining the linear tensor and spherical tensor encoded data in a single b-shell. In particular, we fit a model with a single, "average" compartment, however in reality tissue has been shown to contain multiple compartments with very different diffusion properties.

To test if the presence of multiple compartments would bias the fibre dispersion estimate we model data for tissue with two compartments: an "intra-axonal" compartment with $d_\parallel = 1.7\ \mu m^2/ms$, $d_\perp = 0\ \mu m^2/ms$ (FA=1) and an "extra-axonal" compartment with either the same $d_{iso}$ ($d_\parallel = 1.1\ \mu m^2/ms$, $d_\perp = 0.3\ \mu m^2/ms$, FA=0.68) or a higher $d_{iso}$ ($d_\parallel = 1.7\ \mu m^2/ms$, $d_\perp = 0.9\ \mu m^2/ms$, FA=0.38). In both cases the "extra-axonal" compartment has the same microscopic anisotropy ($d_\parallel - d_\perp = 0.8\ \mu m^2/ms$). We assume both compartments have the same fODF, so they have the same average orientation and the same dispersion of 40° along the major axis and 20° along the minor axis. While this assumption of identical dispersion is likely an oversimplification, it allows us to investigate whether the reconstructed dispersion matches a single "true" dispersion value.

To simulate different types of tissue we vary the "intra-axonal" signal fraction from 0 to 1 with the remaining signal fraction being taken up by the "extra-axonal" compartment (either with the same or different $d_{iso}$). At a signal fraction of 0 or 1 we only have a single compartment and hence the dispersing zeppelin model should be accurate. At intermediate signal fractions we expect our model to break down as we have two compartments with different microscopic anisotropy contributing to the signal. For each signal fraction and "extra-axonal" $d_{\text{iso}}$ we simulate data for 62 volumes acquired using three different acquisition schemes

1. single-shell linear tensor encoding at $b = 1.5$ ms/μm$^2$ for 62 gradient orientations
2. two-shell linear tensor encoding including shells with a b-values of 1.5, and 3 ms/μm$^2$ for 31 gradient orientations each
3. single-shell linear tensor and spherical tensor encoding at $b = 1.5$ ms/μm$^2$. Linear tensor encoding was simulated for 50 gradient orientations and the same spherical tensor encoding was acquired 12 times.

To test both the accuracy and precision of the best-fit parameters in each scenario we simulate 500 noise realizations by adding Rician noise with a standard deviation of 0.033 $S_0$ to each volume (corresponding to an SNR of 30 for the b=0 images). The number of acquisitions and the noise level have been set to resemble the in-vivo data, where the SNR has been estimated from the B0 data with the short echo and repetition time. The SNR is in line with that found by (Szczepankiewicz et al., 2019a).

In practice, the higher b-values in the second scheme or the spherical tensor encoding in the third scheme will require longer diffusion encoding and hence echo time, which would lead to a lower SNR for these acquisitions. While this will bias the estimates of the precision expected for these different acquisition schemes, here we are mainly interested in the accuracy of the fibre dispersion, which should be unaffected.

### *In-vivo data*

For two subjects we acquired linear and spherical tensor encoded data at an isotropic resolution of 2 mm on a 3T Siemens Prisma scanner (192 mm FOV; 6/8 partial Fourier; GRAPPA acceleration of 2; reconstruction using root sum of squares). We gathered 25 axial slices including the full corpus callosum and much of the cortex (covering about half of the subject's brain). To investigate the dependence of the extracted fibre dispersion on the acquisition parameters, we independently vary the b-value (by varying the gradient strength), the repetition time, the echo time, and the gradient duration (see Table 1).Protocols B and D (Table 1) were skipped for one of the subjects due to time constraints. For the spherical tensor encoding 12 volumes were collected per b-value; for the linear tensor encoding 40 volumes with b=1.5 and 60 volumes with b=3 were collected. These were interspersed with b=0 volumes. The total scan time took 45 minutes for all four protocols.

Spherical tensor encoding was acquired using a prototype spin-echo sequence that enables b-tensor encoding (Szczepankiewicz et al., 2019a). The adopted gradient waveform (Figure 1) was numerically optimized using the NOW toolbox in Matlab[8] (Sjölund et al., 2015) and compensated for concomitant gradients (Szczepankiewicz et al., 2019b). During this optimisation, the maximum gradient amplitude and slew rate were set to of the Prisma scanner (i.e, respectively 80 mT/m and 200 mT/m/ms). In practice this maximum gradient amplitude and slew rate were not reached (Figure S1).

The linear tensor encoding data was acquired using a gradient waveform optimised for linear tensor encoding using the NOW toolbox. This gradient waveform is sensitive to longer diffusion times than the ones in the spherical tensor encoding (Figure S1), which might bias the estimate of the microscopic anisotropy if the signal has a strong diffusion time dependence (de Swiet and Mitra, 1996; Jespersen et al., 2019). While for the relatively long diffusion times probed here such a time-dependence of the signal has been found to be small (Clark et al., 2001), using a linear waveform with a matched diffusion time to the spherical tensor encoding would be more accurate (Lundell et al., 2017; Szczepankiewicz et al., 2019a).

---

[8] https://github.com/jsjol/NOW

Table 1 Shells for which both linear tensor and spherical tensor encoded data were acquired for a single subject. The gradient duration is defined as the time from the start to the end of the gradient waveforms (Figure 1). For a given protocol, TE and TR are matched across b-shells and for linear and spherical encoding.

| ID | b-value (ms/µm$^2$) | Repetition time (s) | Echo time (ms) | Gradient duration (ms) |
| --- | --- | --- | --- | --- |
| A | 1.5 & 3.0 | 3.8 | 100 | 72 |
| B | 1.5 & 3.0 | 5.2 | 100 | 72 |
| C | 1.5 & 3.0 | 5.2 | 150 | 72 |
| D | 1.5 & 3.0 | 5.2 | 150 | 120 |

The diffusion data were corrected for motion and distortions using FSL's topup (Andersson et al., 2003) and eddy (Andersson and Sotiropoulos, 2016) tools. When correcting for the distortions in eddy, data from both the b=1.5 and 3 ms/µm$^2$ shells were combined, however eddy was run separately for each repetition time, echo time, diffusion time, as well as separately for the linear tensor and spherical tensor encoded data (for a total of 8 runs). The resulting distortion-corrected, partial-brain data was then registered using a rigid-body transformation to a full-brain b=0 scan acquired at the same time (distortion-corrected using FSL's topup) and finally to a T1-weighted structural scan of the same subject using boundary-based registration (Greve and Fischl, 2009; Jenkinson et al., 2002; Jenkinson and Smith, 2001).

## *Model Fitting*

We fit a single-compartment model of dispersing fibres (eq. 13) to both the multi-compartment simulated data and the in-vivo data. The optimisation was run using the quasi-Newton method L-BFGS-B (Byrd et al., 1995; Zhu et al., 1997). We adopt a Rician noise model to fit the simulations and the in-vivo data.

For the in-vivo data the diffusion data in all shells is fitted simultaneously. During this fit, we only optimise a single set of parameters describing the orientation of the Bingham matrix, which enforces the same mean fibre orientation across all shells. However, the microscopic anisotropy ($d_\parallel - d_\perp$), isotropic diffusion-weighted signal amplitude ($S_{\text{dw}} = S_0 e^{-b\, d_{iso}}$), and fibre dispersion ($k_1$ and $k_2$) are allowed to be different in every shell. This ensures that when we compare the fibre dispersion across different acquisitions, we compare the dispersion around the same mean fibre orientation.

To increase the speed of convergence we iterate between fitting only the three orientation parameters on the full dataset and fitting the other parameters on a per-shell basis. Robustness is increased by initializing microscopic anisotropy and fibre dispersion of each shell using their median value across all shells when fitting them to the shell's diffusion data.

For comparison we also fit NODDI (Zhang et al., 2012) and the ball-and-racket model (Sotiropoulos et al., 2012) to our in-vivo data. In both models we assume the fODF is described by a single Bingham distribution (Tariq et al., 2016) in line with our dispersing zeppelin model. Both models were fitted on GPU using cuDIMOT (Hernandez-Fernandez et al., 2018).

## Results

### Simulations

We investigate the bias incurred in the fibre dispersion estimate when fitting a single-compartment model (i.e. eqs. 4 and 13) to data simulated from a two-compartment tissue with varying "intra-axonal" signal fractions.

Irrespective of how the data was generated, the model is degenerate between fibre dispersion and microscopic anisotropy for a single shell data acquired with linear tensor encoding (left column in Figure 3). The same single-shell diffusion data can be explained by a high microscopic anisotropy and dispersion or a small microscopic anisotropy and dispersion.

Multiple shells obtained with linear tensor encoding (i.e., conventional multi-shell) breaks this degeneracy (middle column in Figure 3). However, the single-compartment model assumes that microscopic anisotropy remains constant across b-values, which is invalid for this data generated from two compartments (except for signal fractions of 0 or 1). This leads to biases in the estimated microscopic anisotropy and hence the mean dispersion. This bias is only a few degrees if both compartments have the same $d_{iso}$ as this ensures that the relative contribution of both compartments to the microscopic anisotropy remains the same across b-values (eqs. 6 and 7). However, if the compartments have very different $d_{iso}$ the microscopic anisotropy at low b-values will be dominated by a different compartment than at high b-values, which leads to a strong dependence of the averaged microscopic anisotropy on b-value. This breaks our assumption of a constant microscopic anisotropy, which leads to a large bias in the fibre dispersion (Figure 3H).

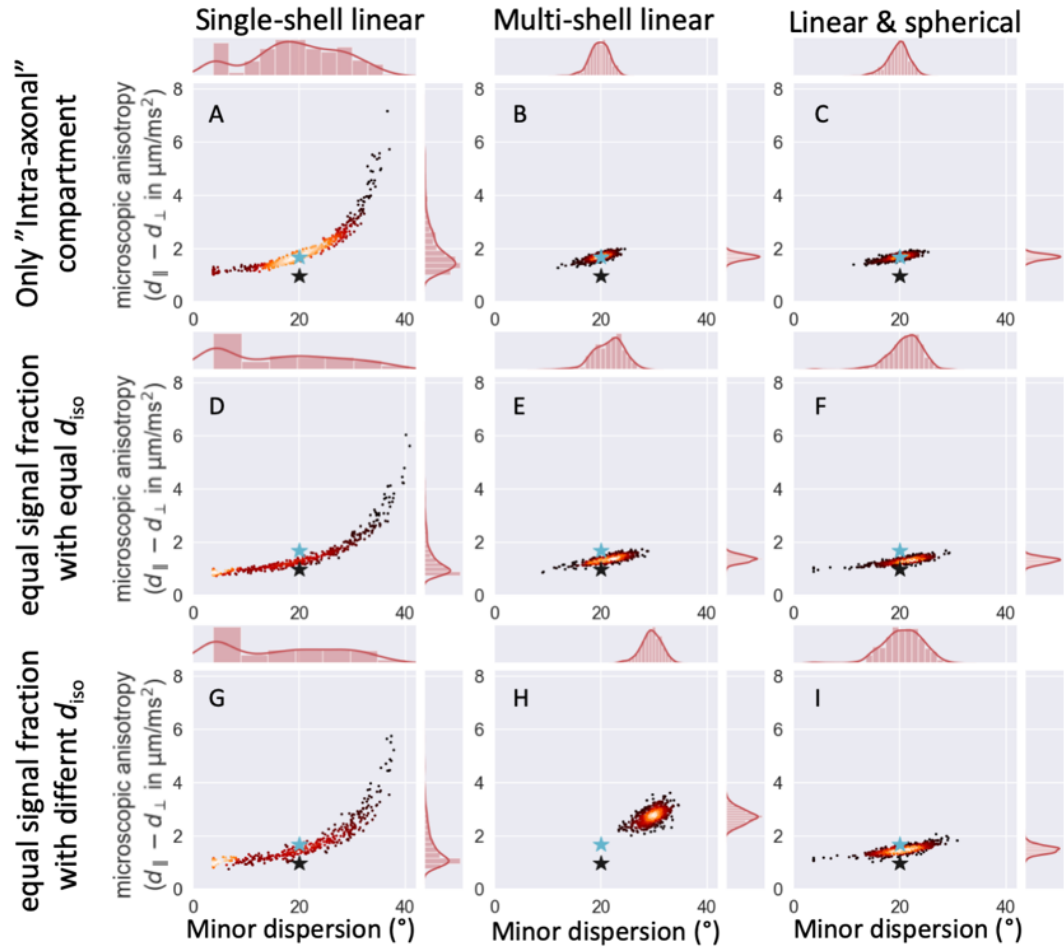

**Figure 3** Best-fit parameter estimates for 500 noise realizations using three different acquisition schemes (from left to right: single-shell linear tensor encoding, multi-shell linear tensor encoding, and single-shell linear and spherical tensor encoding) for three different underlying anatomies, namely only an "intra-axonal" compartment (top), equal signal fraction for an "intra-axonal" and "extra-axonal" compartment with equal $d_{iso}$ (middle) or different $d_{iso}$ (bottom). In all cases the microscopic anisotropies for the two compartments are different, but the dispersion is the same (marked by cyan star for "intra-axonal" and black star for "extra-axonal"). Either adding multiple shells (middle column) or adding spherical tensor encoded data (right column) breaks the degeneracy seen between the parameters in the single-shell linear tensor encoding (left column), however only for the addition of spherical tensor encoded data does this not lead to a bias if the multiple compartments are not modelled correctly.

The addition of spherical tensor encoding gives an accurate estimate of a weighted average of the microscopic anisotropy, which equals that of the "intra-axonal" compartment if that is the only compartment present (Figure 3C) or the average of the "intra-axonal" and "extra-axonal" microscopic anisotropies if both are present (Figure 3F,I). Because this estimate of the microscopic anisotropy is obtained from a single shell of diffusion data, it is unaffected by our model assumption of the dependence of the microscopic anisotropy on b-value. That this weighted average of the microscopic anisotropy provides an accurate estimate of the width of the response function is illustrated by the reduced bias in the estimate of fibre dispersion (Figure 3I compared with H). See Figure S2-S4 for the correlations between all parameter estimates.

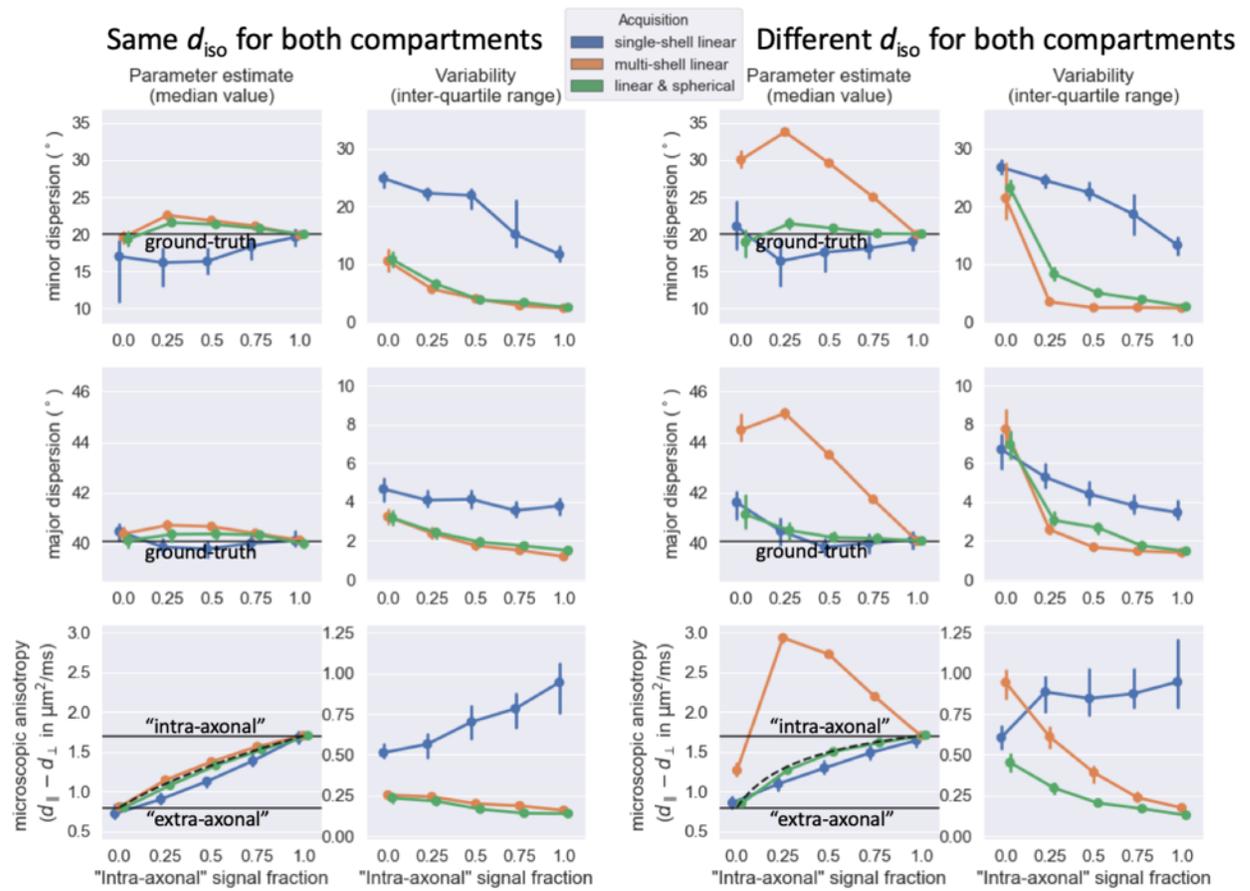

**Figure 4** For simulations where the "intra-axonal" and "extra-axonal" compartments have the same isotropic diffusivity (left) or different isotropic diffusivity (right), the median (left) and interquartile range (right) estimated from 500 noise iterations using three different acquisition schemes (color-coded according to the legend at the top). Black lines

mark the ground truth. In the lower panels the black dashed lines illustrate the expected micro-anisotropy when approximated as a weighted average (eq. 9). Error bars indicate 95% confidence intervals of the median and inter-quartile range estimated by bootstrapping the 500 simulations. Even though the different acquisitions schemes would in practice have different echo times, for simplicity we assumed the same number of volumes (i.e., 62) and SNR (i.e., 30) for all acquisitions.

Figure 4 illustrates the accuracy and variability of the estimates for the full range of simulations. For single-shell linear encoding (in blue) the degeneracy between the microscopic anisotropy and dispersion leads to a large variability between the noise realisations. While both multi-shell diffusion data (in orange) or the inclusion of spherical tensor encoding (in green) break the degeneracy, the values from the multi-shell data are only accurate if the model assumption of no dependence of the microscopic anisotropy on b-value is accurate (i.e., if there is only a single compartment or if the multiple compartments have the same $d_{\text{iso}}$). Irrespective of the $d_{\text{iso}}$ the microscopic anisotropy smoothly increases from the "extra-axonal" to the "intra-axonal" microscopic anisotropy as the signal fraction of the "intra-axonal" compartment increases for data including spherical tensor encoding (green in Figure 4) in line with the trend expected from computing the micro-anisotropy as a weighted average (eq. 9; black dashed line in Figure 4). This more realistic trajectory of estimated microscopic anisotropy reduces the bias in the fibre dispersion in the case of multiple compartments with different $d_{\text{iso}}$, although it is not fully eliminated.

In our in-vivo scans we cannot manipulate the intra-axonal volume fraction; however, we can change the relative contribution of the tissue compartments by altering the acquisition parameters. In our simulations, we test this by varying the reference b-value from the value of 1.5 ms/µm² used in Figure 4. When the compartments have the same $d_{\text{iso}}$, altering the b-value does not change the relative contribution of the compartments, which leads to a constant microscopic anisotropy and fibre dispersion measured across b-values (left in Figure 5). With a lower $d_{\text{iso}}$ for intra-axonal space, increasing the b-value increases the relative contribution of this compartment, leading to an increase in the microscopic anisotropy, although the fibre dispersion still remains nearly constant as long as spherical tensor encoding was included in the acquisition (right in Figure 5).

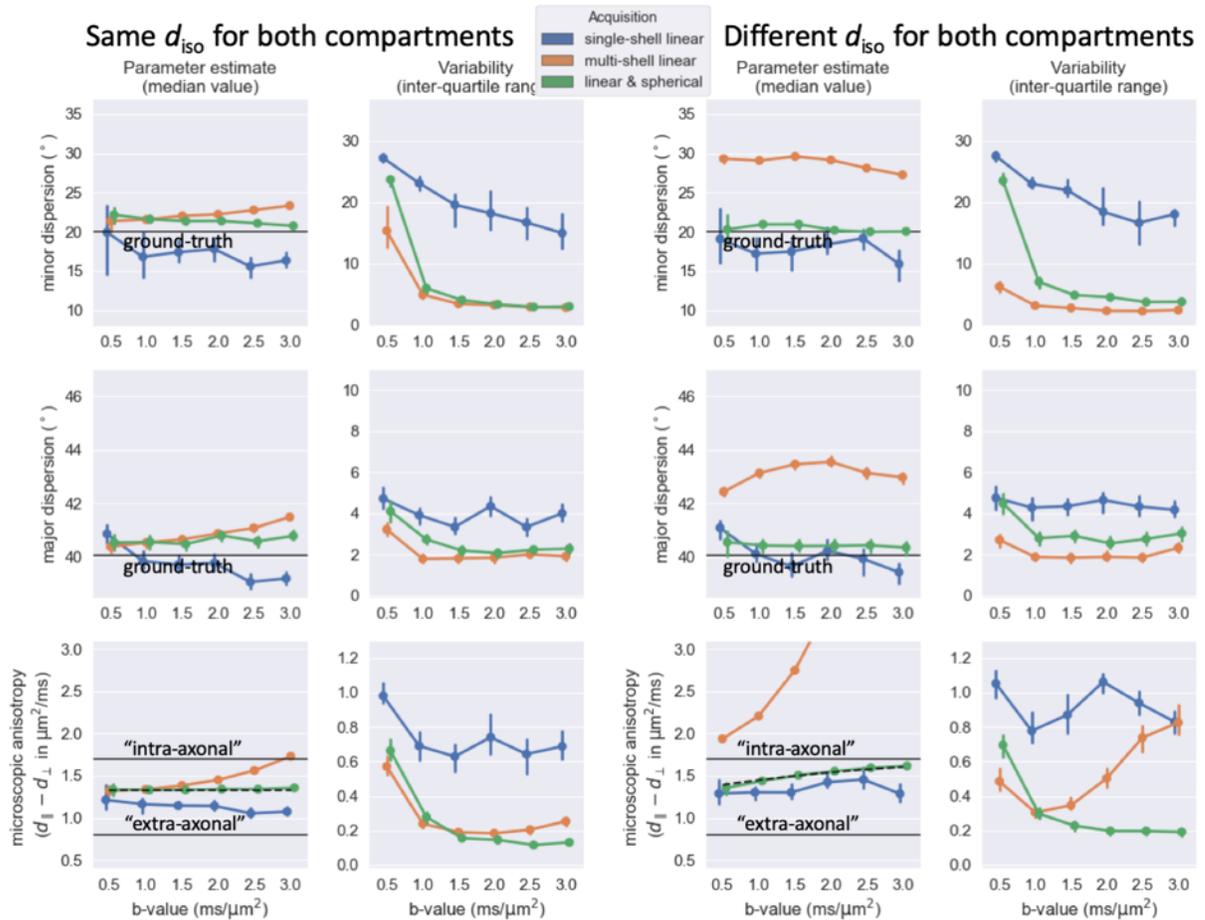

**Figure 5** Similar trend lines as Figure 4, but as a function of the b-value rather than the volume fraction, which is kept fixed at 0.5. The multi-shell data includes two shells with the reference b-value and twice the reference b-value.

## *In-vivo fibre dispersion*

The simulated data above suggests that a single shell of linear tensor and spherical tensor encoded data provides a nearly unbiased measure of fibre dispersion. We cannot confirm this in-vivo due to a lack of a ground truth. However, we can test whether this measure of fibre dispersion remains consistent as the acquisition parameters change. Here we test the sequence for two healthy subjects in-vivo. The result for subject A is shown in Figure 6; for subject B in Figure S5. When available values for both subjects are reported using the following notation <value for subject A>|<value for subject B>.

As a reference, we will adopt the best-fit estimates for a shell with b-value of 3 ms/μm$^2$, TR=3.8s, TE=100ms, and a total duration of the gradient waveform of 72 ms (A in Table 1). The microscopic anisotropy is highest in the white matter (median of 1.7|1.7 μm$^2$/ms) with no strong decrease in crossing-fibre regions as seen for a fractional anisotropy map (Figure 6). While the microscopic anisotropy in the grey matter is lower than in the white matter (median of 1.1|1.1 μm$^2$/ms), this is still a much smaller difference than the near isotropic diffusion typically seen in cortical grey matter in fractional anisotropy maps.

Because our model does not explicitly allow for crossing fibres, the major axis of dispersion tends to be oriented along the plane of the crossing fibres with high dispersion values and is close to the maximum of 60° for those regions with crossing fibres. The dispersion along the minor axis reflects the dispersion along an axis perpendicular to the crossing fibres and hence is more likely to closely reflect the actual dispersion which ranges from 20-30° in the corpus callosum to ~50° in the centrum semiovale.

At half the reference b-value, the best-fit microscopic anisotropy is increased by 4|3% in white matter and 3|3% grey matter (top row of scatter plots in Figure 6), which corresponds to a decrease of about 48% in the width of the response function (i.e., *b*-value multiplied with the microscopic anisotropy). However, the signal anisotropy between gradient orientations also greatly decreases as the b-value is decreased, which leads to a net shift in the fibre dispersion estimates of on average 1°|1.2° (which is in line with the minimal changes seen in the simulations; green in Figure 5).

We explored a range of other acquisition parameters to investigate whether they introduced a bias in the fibre dispersion estimates. Increasing the repetition time from 3.8 to 5.2 s has little effect on the signal attenuation and hence the best-fit parameters (second row of scatter plots in Figure 6). When the echo time is also increased from 100 to 150 ms, we find a 4%|3% decrease in the microscopic anisotropy and a systematic increase in the fibre dispersion estimated in the white matter (~1.2|0.8°) and gray matter (~1.4|1.7°).

When the gradient duration is increased by 60% (from 72 to 115 ms) this causes a further decrease in the microscopic anisotropy to a total of 6% in white matter and 7% in gray matter. This does not appear to significantly affect the fibre dispersion which increases by ~1.1° in white matter and ~1.3° in gray matter.

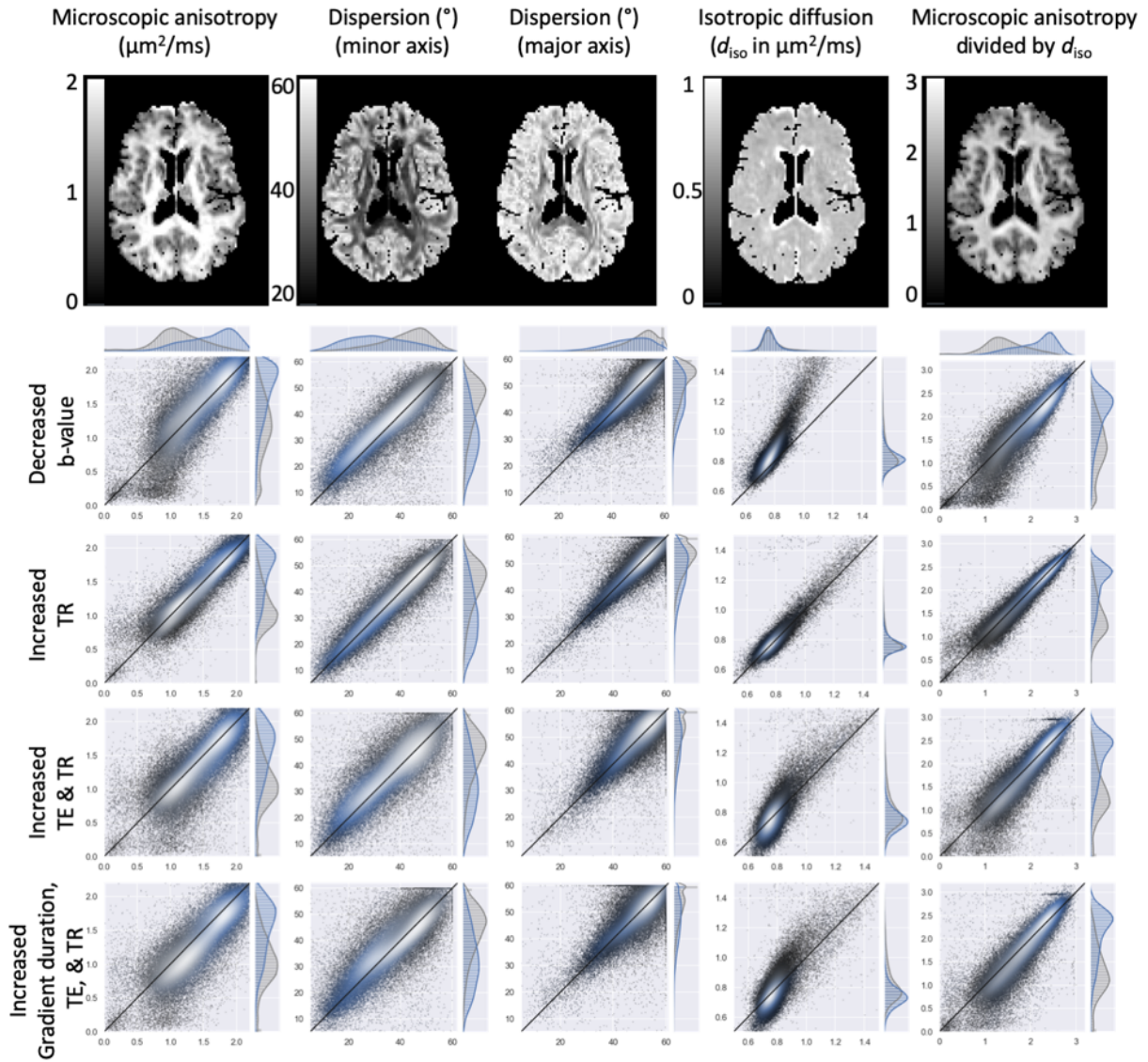

**Figure 6** Consistency of from left to right the best-fit microscopic anisotropy, fibre dispersion (along both minor and major axis), the isotropic diffusivity, and the microscopic anisotropy normalised by the isotropic diffusivity compared between different acquisitions of the same subject. The top row shows an axial slice for data acquired with b=3 ms/µm², TR=3.8 s, TE=100 ms and a short gradient duration. The subsequent rows compare these fits (on the x-axis) for all white matter (blue) and gray matter (gray) voxels with those acquired for a decreased b-value (1.5 ms/µm²), an increased TR (to 5.2 s), increased TR and TE (to 150 ms) and an increased TR, TE, and gradient duration (which effectively increases the diffusion time). Lighter colors indicate a higher density of points.

Finally, we note that the increase in the microscopic anisotropy for lower b-value corresponds to a similar increase in the isotropic diffusivity (fourth column in Figure 6). In other words, the microscopic anisotropy normalized by the isotropic diffusivity (last column in Figure 6) does not depend on b-value, although it does change with the microscopic anisotropy when the diffusion time changes. The isotropic diffusivity was estimated from the mean spherical tensor encoded attenuation using eq. 4.

The fibre dispersion estimated from the dispersing zeppelin model that includes information from the spherical tensor encoded signal are systematically lower (on average 0.9|1.3° in white matter, 1.9|7.8° in grey matter) than those estimated from the ball-and-racket model (Figure 7). Compared with NODDI, the fibre dispersions are higher in the grey matter (2.5|2.6° on average), but lower in the white matter (0.7|0.4° on average).

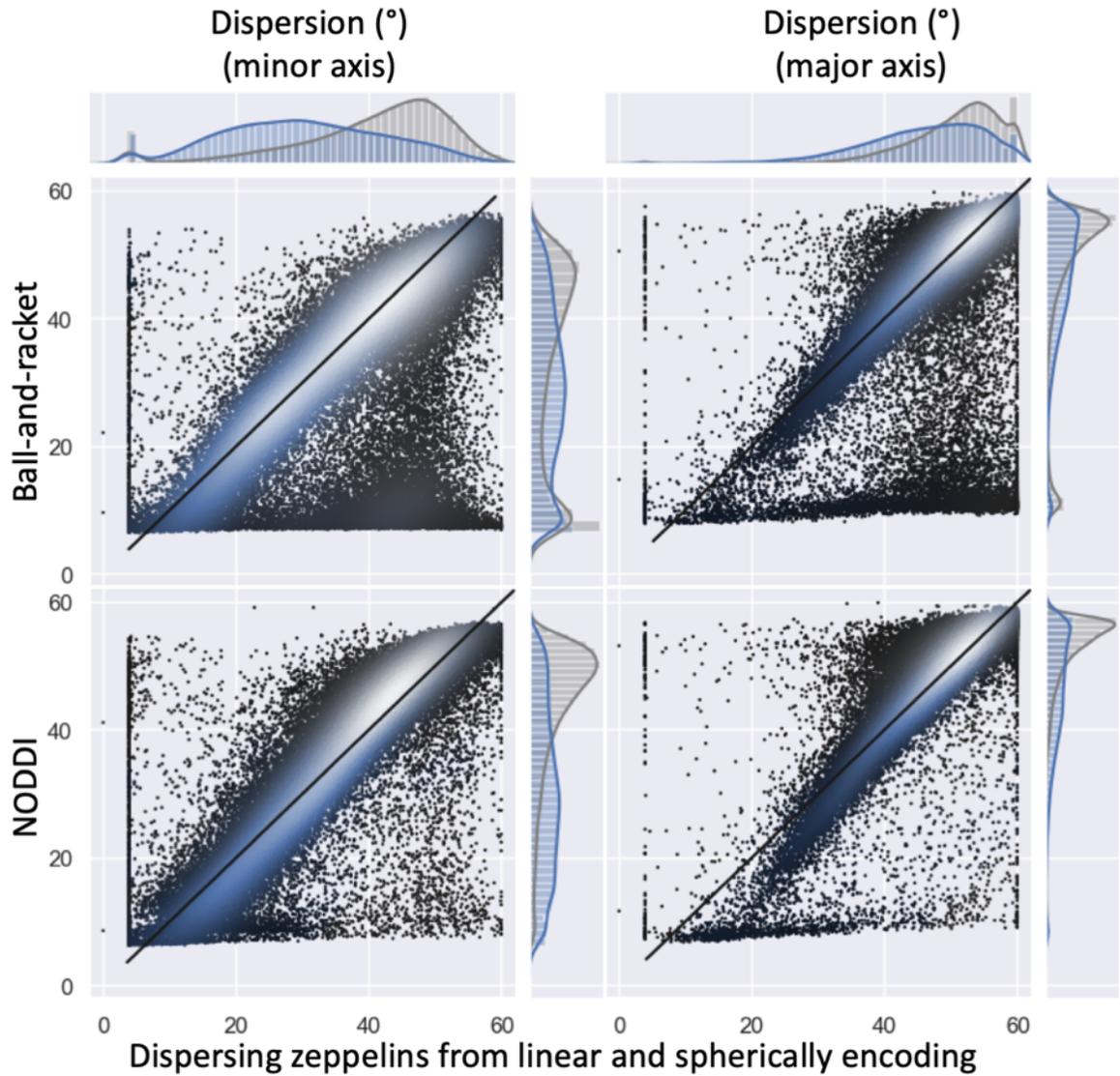

Figure 7 Fibre dispersion estimates (in degrees) compared between the dispersing zeppelin model constrained by the spherical tensor encoding data on the x-axis (for b=3 ms/µm$^2$) with NODDI (Tariq et al., 2016; Zhang et al., 2012) and the ball-and-racket model (Sotiropoulos et al., 2012) (for both b-values). The left column shows fibre dispersion along the minor axis; the right column along the major axis. Like in Figure 6 in the scatter plots and histograms the white matter voxels have been represented in blue and the grey matter voxels in grey.

## Discussion

Here we argued that microscopic anisotropy obtained by combining linear and spherical tensor encoding allows us to deconvolve the diffusion signal to obtain an accurate measure of fibre dispersion from single shell diffusion data. Although both microscopic anisotropy and fibre response function can be expressed using the difference between axial and radial diffusivities ($d_\parallel - d_\perp$) (Dell'Acqua et al., 2007), this result is not trivial since realistic tissue cannot be described by a single diffusion anisotropy ($d_\parallel - d_\perp$), but is likely to contain many compartments each with different diffusion properties.

In our simulations of multi-compartment tissue, we find that the combination of linear and spherical tensor encoding produces more accurate estimates of the fibre dispersion than multi-shell data even while we fit a single-compartment model to multi-compartment data (Figure 4) with systematic biases remaining of up to 2-3°. We speculate that this is because the ratio of the linear tensor and spherical tensor encoded signals for multiple compartments (eq. 6) produces an estimate of the microscopic anisotropy that is approximately the average of the microscopic anisotropy of the individual components weighted by the component's signal fraction (eq. 9). This additive nature of the microscopic anisotropy in the second-order signal expansion was previously noted by (Jespersen et al., 2013). This approximation is expected to hold up to up to $b(d_\parallel - d_\perp) \approx 6$ (Figure 2). During the averaging of the microscopic anisotropy each compartment is weighted by $S_0 e^{-b\, d_{\text{iso}}}$ (eq. 7). This b-value dependent weighted average breaks the degeneracy between the microscopic anisotropy and fibre dispersion inherent in the linear tensor encoding at the appropriate fibre dispersion (Figure 3C,I).

The accuracy of the fibre dispersion estimate suggested by these simulations can be tested by investigating the consistency of the fibre dispersion in the in-vivo data when varying the acquisition parameters. As the b-value, echo time, and gradient duration are varied we found changes in the microscopic anisotropy changes of up to 7% and a systematic offset in the dispersion of typically 1-2° (Figure 6). This is in line with the systematic bias in fibre dispersion remaining in the simulations for linear and spherical encoded data. Hence, these small variations in fibre dispersion for different acquisition parameters are consistent with the single-compartment model giving an accurate measure of a "true" fibre dispersion up to an accuracy of a few degrees. However, without direct comparison to a ground-truth fibre dispersion from histology, the evidence for the increased accuracy remains primarily based on the simulations.

We adopted a single-compartment model, because it makes a simple assumption that the microscopic anisotropy does not depend on b-value. This will hold as long as there only is a single compartment or all compartments have the same isotropic diffusivity and hence the average microscopic anisotropy does not depend on b-value (eqs. 6 and 7). In those situations, the microscopic anisotropy estimates are expected to be the same from multi-shell linear tensor encoding or single-shell linear and spherical tensor encoding, and both acquisitions will give the same fibre dispersion estimates (Figure 4). However, as the model assumptions break down (e.g., multiple compartments exist with different isotropic diffusivity) the bias in the fibre dispersion increases for multi-shell data (Figure 4).

Even if the average microscopic anisotropy is accurately estimated (either from spherical tensor encoding or an accurate model of how the microscopic anisotropy depends on b-value), we still find a systematic bias in the fibre dispersion of about 2-3°. This possibly reflects that a single diffusion tensor even with an appropriately averaged microscopic anisotropy cannot fully capture the angular dependence of the linear tensor encoded data generated from multi-compartment tissue. More accurate microstructural models applied to linear and spherical tensor encoding are likely to further reduce this bias.

Another limitation is that in the simulations we assumed that each compartment had the same dispersion, which allowed us to compare the best-fit fibre dispersion with a single ground-truth value. However, in reality different compartments might have different orientational distributions. For example, small axons might have a different fODF than large axons and both of these might differ from the ODF of the extra-axonal compartment. In that case there would be no single ground-truth fibre dispersion value. If the fibre dispersions are very different in different compartments, one might expect the estimated fibre dispersion to change when changing the acquisition parameters, which alters the relative sensitivity to the different compartments. We find no evidence for that in the in vivo data.

While the fibre dispersion from multiple b-shell models have been shown to correlate with the fibre dispersion measured using microscopy in post-mortem tissue, potential systematic offsets between the diffusion MRI and microscopy estimates remained (Mollink et al., 2017). In the in-vivo data we found offsets of ~1-3° on average between the fibre dispersion estimates from our model including spherical tensor encoding and those from the ball-and-racket model (Sotiropoulos et al., 2012) and NODDI (Tariq et al., 2016; Zhang et al., 2012). These offsets might reflect the bias found in the simulations when fitting our single-compartment model to multi-compartment data. So, in practice we don't find the large deviations of several 10s of degrees, which our simulations suggest are possible between fibre dispersion estimates from the multi-shell diffusion data or the linear and spherical tensor encoded diffusion data. This suggests that the microstructural models in these models are at least in the healthy brain accurate enough to get reliable fibre dispersion estimates. This reliability is expected to go down when the isotropic diffusivities between compartments becomes more different (Figure 4) as expected in some pathologies, such as white matter lesions (Lampinen et al., 2019).

The fibre dispersion can also be estimated without spherical tensor encoding or multi-shell data by assuming a constant response function throughout the white matter (Tournier et al., 2004). Recent reports have cast doubt on the validity of this assumption of a constant response function even in healthy white matter (Schilling et al., 2018; Howard et al., 2019), which is likely to be worse in disease states. However, whether the biases found in the response function are large enough to significantly bias the fibre dispersion estimates, should still be investigated.

We conclude that fibre dispersion estimated from multiple b-values are more sensitive to the assumptions made about the microstructural tissue parameters than the fibre dispersions estimated from a single b-shell with linear tensor and spherical tensor encoded data.

## *Acknowledgements*

MC and SS were partially supported by the EPSRC UK (EP/L023067). SJ was supported by the MRC UK (Grant Ref: MR/L009013/1). MN is supported by the CR Award (MN15), the Swedish Research Council (grant no. 2016-03443). The Wellcome Centre for Integrative Neuroimaging is supported by core funding from the Wellcome Trust (203139/Z/16/Z). We thank Jon Campbell for help in acquiring the data and Jelle Veraart and Sune Jespersen for helpful discussions.

## *Declaration of interest*

MN declares research support from and ownership interest in Random Walk Imaging (formerly Colloidal Resource), and patent applications in Sweden (1250453-6 and 1250452-8), USA (61/642 594 and 61/642 589), and PCT (SE2013/ 050492 and SE2013/050493). FS has been employed at Random Walk Imaging. Remaining authors declare no conflict of interest.

## *References*

Anderson, A.W., 2005. Measurement of fiber orientation distributions using high angular resolution diffusion imaging. Magn Reson Med 54, 1194–206. https://doi.org/10.1002/mrm.20667

Andersson, J.L., Skare, S., Ashburner, J., 2003. How to correct susceptibility distortions in spin-echo echo-planar images: application to diffusion tensor imaging. Neuroimage 20, 870–88. https://doi.org/10.1016/S1053-8119(03)00336-7

Andersson, J.L., Sotiropoulos, S.N., 2016. An integrated approach to correction for off-resonance effects and subject movement in diffusion MR imaging. Neuroimage 125, 1063–78. https://doi.org/10.1016/j.neuroimage.2015.10.019

Basser, P.J., Mattiello, J., LeBihan, D., 1994. MR diffusion tensor spectroscopy and imaging. Biophysical Journal 66, 259–267. https://doi.org/10.1016/s0006-3495(94)80775-1

Basser, P.J., Pajevic, S., Pierpaoli, C., Duda, J., Aldroubi, A., 2000. In vivo fiber tractography using DT-MRI data. Magn Reson Med 44, 625–32.

Behrens, T.E., Berg, H.J., Jbabdi, S., Rushworth, M.F., Woolrich, M.W., 2007. Probabilistic diffusion tractography with multiple fibre orientations: What can we gain? Neuroimage 34, 144–55. https://doi.org/10.1016/j.neuroimage.2006.09.018

Byrd, R.H., Lu, P., Nocedal, J., Zhu, C., 1995. A limited memory algorithm for bound constrained optimization. SIAM Journal on Scientific Computing 16, 1190–1208.

Callaghan, P.T., Jolley, K.W., Lelievre, J., 1979. Diffusion of water in the endosperm tissue of wheat grains as studied by pulsed field gradient nuclear magnetic resonance. Biophysical Journal 28, 133–141. https://doi.org/10.1016/s0006-3495(79)85164-4

Clark, C.A., Hedehus, M., Moseley, M.E., 2001. Diffusion time dependence of the apparent diffusion tensor in healthy human brain and white matter disease. Magn Reson Med 45, 1126–9.

Coelho, S., Pozo, J.M., Jespersen, S.N., Jones, D.K., Frangi, A.F., 2019. Resolving degeneracy in diffusion MRI biophysical model parameter estimation using double diffusion encoding. Magn Reson Med 82, 395–410. https://doi.org/10.1002/mrm.27714

de Swiet, T., Mitra, P., 1996. Possible Systematic Errors in Single-Shot Measurements of the Trace of the Diffusion Tensor. J Magn Reson B 111, 15–22. https://doi.org/10.1006/jmrb.1996.0055

Dell'Acqua, F., Rizzo, G., Scifo, P., Clarke, R.A., Scotti, G., Fazio, F., 2007. A model-based deconvolution approach to solve fiber crossing in diffusion-weighted MR imaging. IEEE Trans Biomed Eng 54, 462–72. https://doi.org/10.1109/TBME.2006.888830

Dell'Acqua, F., Scifo, P., Rizzo, G., Catani, M., Simmons, A., Scotti, G., Fazio, F., 2010. A modified damped Richardson-Lucy algorithm to reduce isotropic background effects in spherical deconvolution. Neuroimage 49, 1446–58. https://doi.org/10.1016/j.neuroimage.2009.09.033

Dell'Acqua, F., Tournier, J.D., 2018. Modelling white matter with spherical deconvolution: How and why? NMR Biomed e3945. https://doi.org/10.1002/nbm.3945

Descoteaux, M., Angelino, E., Fitzgibbons, S., Deriche, R., 2007. Regularized, fast, and robust analytical Q-ball imaging. Magn Reson Med 58, 497–510. https://doi.org/10.1002/mrm.21277

Eriksson, S., Lasic, S., Topgaard, D., 2013. Isotropic diffusion weighting in PGSE NMR by magic-angle spinning of the q-vector. J. Magn. Reson. 226, 13–8. https://doi.org/10.1016/j.jmr.2012.10.015

Greve, D.N., Fischl, B., 2009. Accurate and robust brain image alignment using boundary-based registration. Neuroimage 48, 63–72. https://doi.org/10.1016/j.neuroimage.2009.06.060

Hernandez-Fernandez, M., Reguly, I., Jbabdi, S., Giles, M., Smith, S., Sotiropoulos, S.N., 2018. Using GPUs to accelerate computational diffusion MRI: From microstructure estimation to tractography and connectomes. Neuroimage 188, 598–615. https://doi.org/10.1016/j.neuroimage.2018.12.015

## *Supplementary figures*

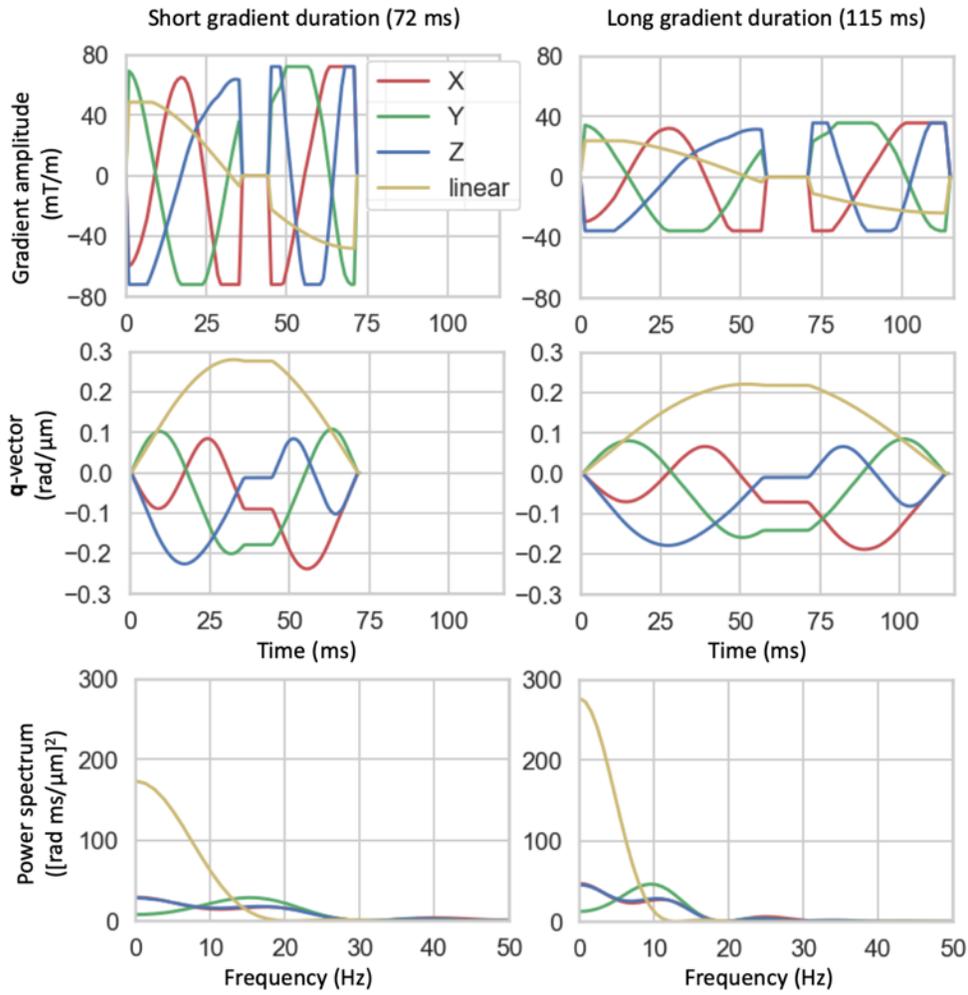

**Figure S1** Gradient waveform (top), **q**-vector (middle), and corresponding power spectrum (bottom) for the spherical tensor encoding (red, green and blue for respectively x-, y-, and z-gradients) and linear tensor encoding (yellow) for the short gradient duration (i.e., A, B, and C in Table 1) and long gradient duration (i.e., D in Table 1). The waveforms are shown for b=3. For b=1.5 the gradients amplitude and **q**-vector are reduced by a factor of $\sqrt{2}$ and the power by a factor of 2.

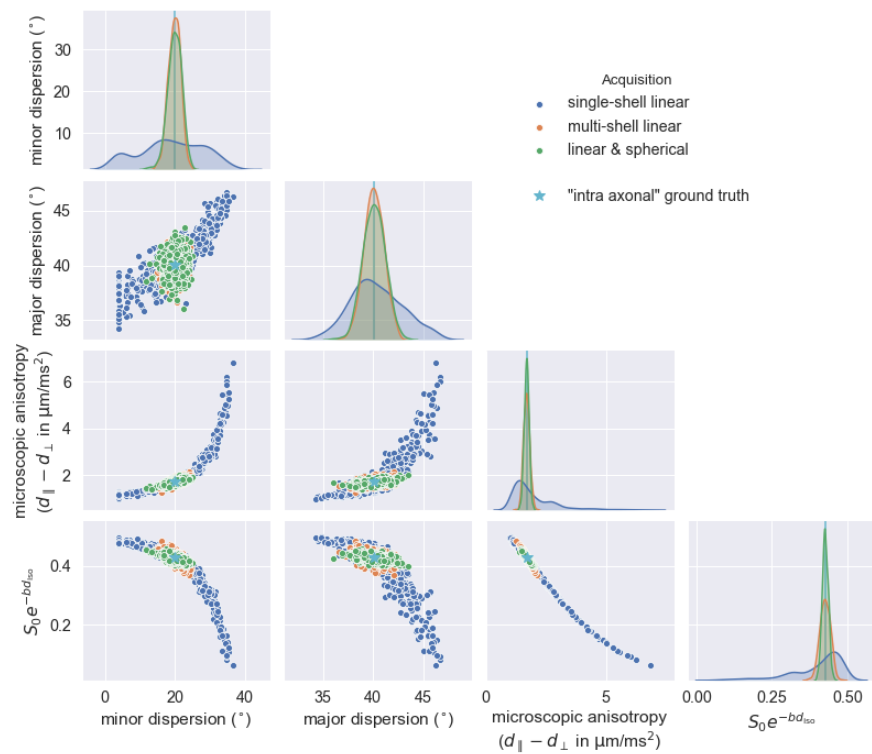

**Figure S2** Correlation and distribution of the best-fit parameter estimates for the 500 noisy realisations of tissue containing only the "intra-axonal" component. From top to bottom and left to right these are the dispersion along the axis with the least dispersion (20°), the dispersion along the axis with the most dispersion (40°), the microscopic anisotropy in µm²/ms, and signal amplitude weighted by the isotropic diffusivity. The stars show the ground-truth values for tissue consisting purely of "intra-axonal" water (cyan) or "extra-axonal "water (black). In this scenario the single-compartment model is valid, so both the multi-shell linear tensor encoding (orange) or linear and spherical tensor encoding (green) give accurate dispersion estimates.

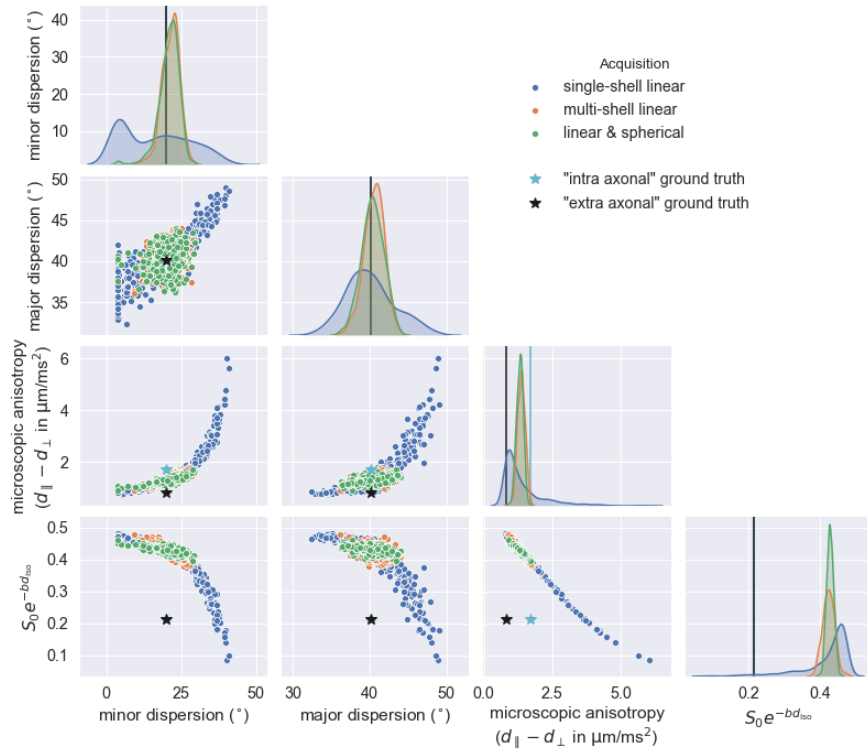

**Figure S3** Same as Figure S2, but for simulated tissue with an "intra-axonal" and "extra-axonal" compartment with the same signal fraction and the same isotropic diffusivities. In this scenario the single-compartment model is invalid, but the assumption of a constant microscopic anisotropy with b-value is valid, so both the multi-shell linear tensor encoding (orange) or linear and spherical tensor encoding (green) have the same bias in the dispersion estimates.

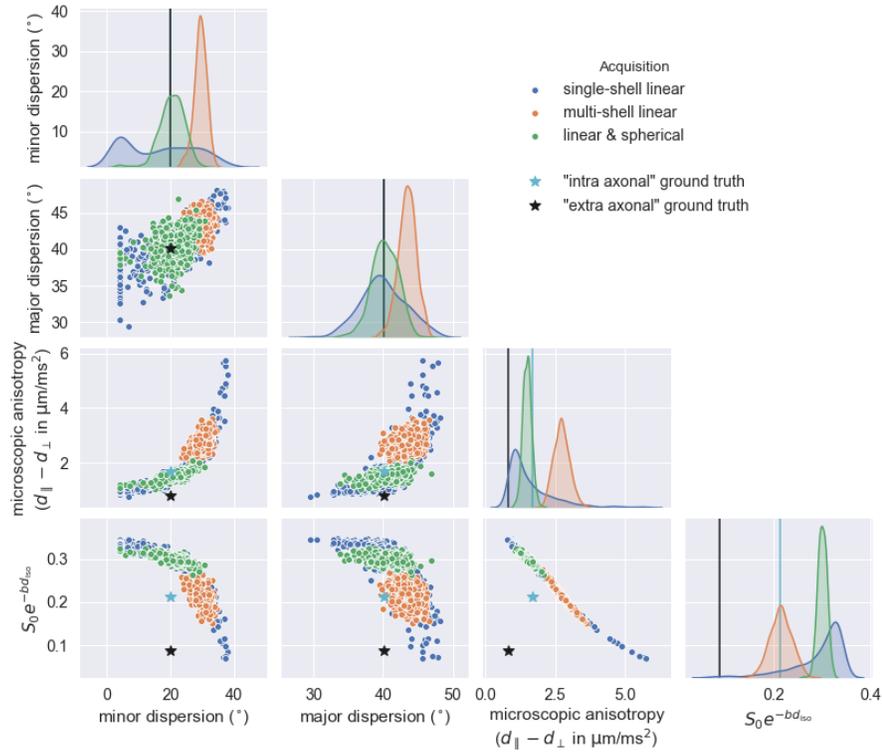

**Figure S4** Same as Figure S2, but for simulated tissue with an "intra-axonal" and "extra-axonal" compartment with the same signal fraction and different isotropic diffusivities. In this scenario the single-compartment model assumption of a constant microscopic anisotropy with b-value is invalid, which leads to a larger systematic bias in the microscopic anisotropy and hence dispersion estimated for the multi-shell linear tensor encoding (in orange) compared with the single-shell linear and spherical tensor encoding (in green).

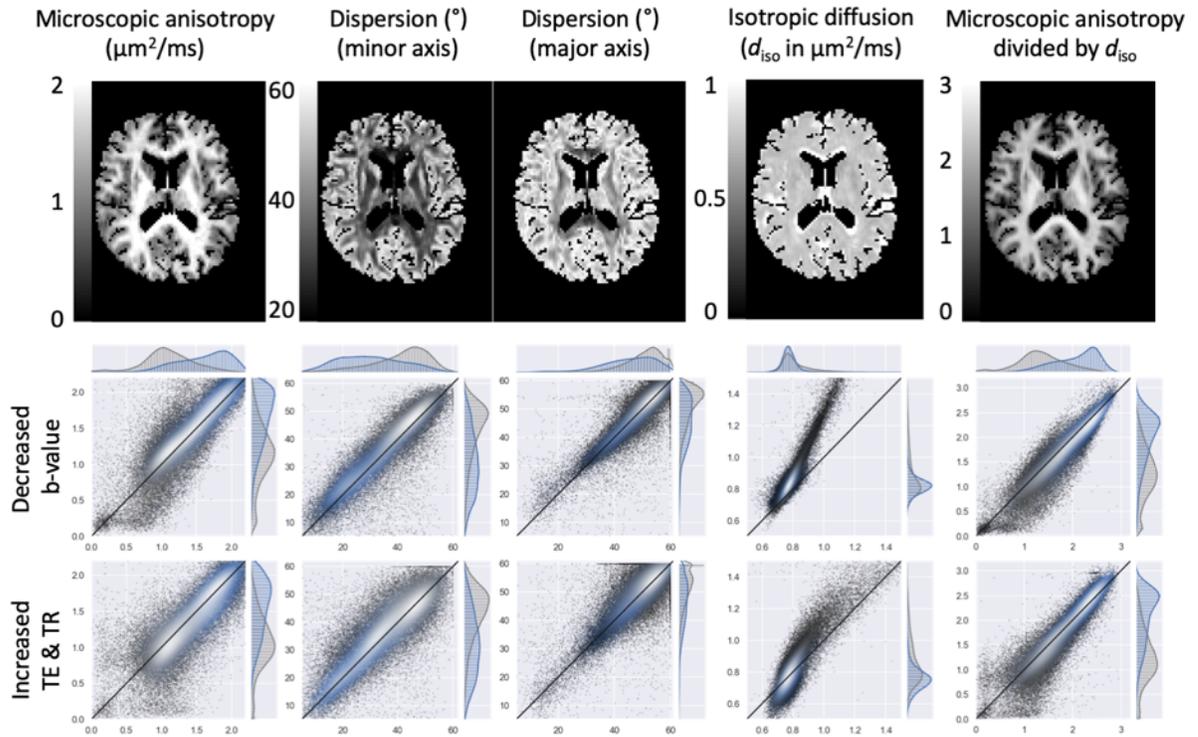

**Figure S5** Equivalent of Figure 6 for a second subject (note that due to time constraints the scans with increased TR, but not increased TE, and the scans with increased gradient duration were skipped for this subject).